\documentclass[review]{elsarticle}

\usepackage{hyperref}
\usepackage{graphicx}
\usepackage{subfigure}
\usepackage{threeparttable}
\usepackage{booktabs}
\usepackage{amstext}

\journal{arXiv}

\begin{document}

\begin{frontmatter}

\title{An Empirical Study on Academic Commentary and Its Implications on Reading and Writing}


\author[WT,WH]{Tai Wang\corref{mycorrespondingauthor}}
\cortext[mycorrespondingauthor]{Corresponding author}
\ead{wangtai@mail.ccnu.edu.cn}

\author[XH,HC,WH]{Xiangen Hu}
\ead{xhu@memphis.edu}

\author[HC]{Keith Shubeck}
\ead{keithshubeck@gmail.com}

\author[HC]{Zhiqiang Cai}
\ead{zcai@memphis.edu}

\author[JT]{Jie Tang}
\ead{jietang@tsinghua.edu.cn}

\address[WT]{National Research \& Engineering Center for E-learning, Wuhan, China}

\address[XH]{Department of Psychology, University of Memphis, TN, United States}

\address[HC]{Institute for Intelligent Systems, University of Memphis, TN, United States}

\address[WH]{Key Lab. of Adolescent Cyberpsychology and Behavior, Ministry of Education, China}

\address[JT]{Department of Computer Science and Technology, Tsinghua University, Peking, China}

\begin{abstract}
The relationship between reading and writing (RRW) is one of the major themes in learning science. One of its obstacles is that it is difficult to define or measure the latent background knowledge of the individual. However, in an academic research setting, scholars are required to explicitly list their background knowledge in the citation sections of their manuscripts. This unique opportunity was taken advantage of to observe RRW, especially in the published academic commentary scenario. RRW was visualized under a proposed topic process model by using a state of the art version of latent Dirichlet allocation (LDA). The empirical study showed that the academic commentary is modulated both by its target paper and the author's background knowledge. Although this conclusion was obtained in a unique environment, we suggest its implications can also shed light on other similar interesting areas, such as dialog and conversation, group discussion, and social media.
\end{abstract}

\begin{keyword}
Read and Write \sep Latent Dirichlet Allocation \sep Background Knowledge \sep Topic Process Model

\end{keyword}

\end{frontmatter}


\section{Introduction}

The relationship between reading and writing (RRW) is a commonly explored theme in the learning sciences and educational psychology \cite{Harl2013} \cite{Abbott2010} \cite{Mateos2011}. Experiments exploring this relationship typically examine the background knowledge of participants, which is commonly assessed by using a pre-test. Unfortunately, it is not easy to construct appropriate questions for these pre-tests. If the pre-test questions are too similar to the experiment's content, participants can extract clues from the pre-test questions, thereby artificially improving their performance on the experiment's task. Alternatively, if the pre-test questions are too dissimilar from the experiment's content, participants can experience interference and perform worse on the experiment's task. In an effort to avoid the issues associated with pre-test bias on background knowledge assessment, experimenters should be encouraged to provide a more natural setting where participants can report their own background knowledge, uninhibited and voluntarily.

A setting ripe for research in the background knowledge and RRW domain that meets the above qualities is that of scholars and their academic papers. Scholars are obligated to cite and append all of the references they believe to be most relevant to their work, due to the ethic codes of academic research communities. These references are usually used as evidence to support their hypotheses, or as sources to establish the motivation behind their studies. Therefore, if we narrow the definition of background knowledge to the foundation which is necessary to build a particular academic paper upon, it is reasonable to assume that these references are the background knowledge of the author of any particular academic paper. Worth noting is the exclusion of basic or common knowledge in this definition of background knowledge. For example, simple arithmetic or basic calculus is not included in the background knowledge of a theoretical physicist when writing an academic paper, because the paper will likely be published in a limited academic circle.

There are roughly two kinds of academic papers. One type of academic paper reports innovations, and the other, academic commentaries, comment on a prior published work. From the perspective of the stimulus-response model, an academic commentary can be seen as the response to the stimulus provided by the original paper. It can also be seen as the output of the author's mind, where the input is the original paper that the author is commenting on.

\begin{figure}[htbp]
\centering   
\includegraphics[width = 6 cm]{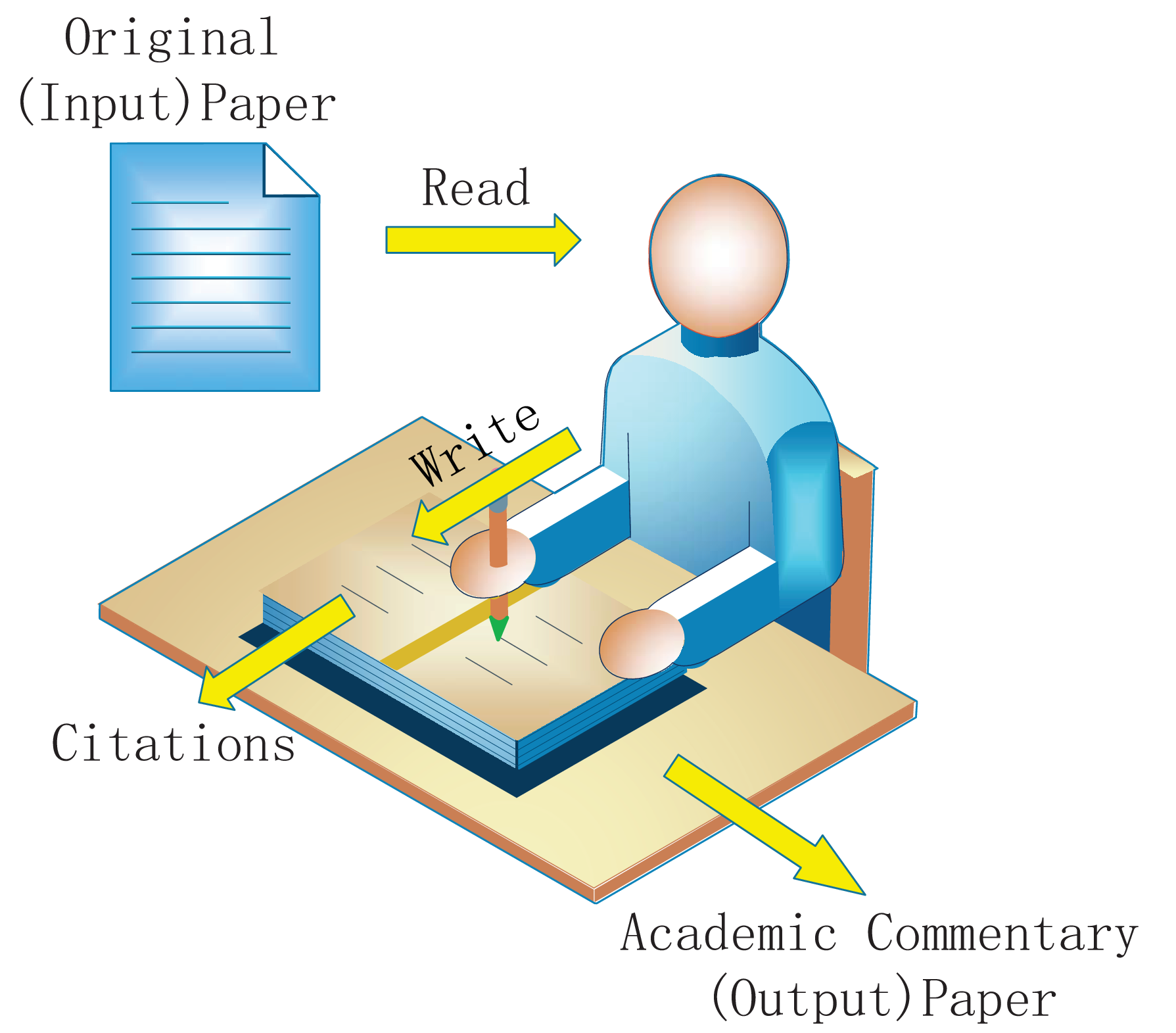}
\caption{Input(Reading) Output(Writing) Relationship in Academic Commentary Scenario} 
\label{fig:1} 
\end{figure}

Fig.\ref{fig:1} shows the academic commentary scenario.

This scenario not only provides a chance to disclose the background knowledge of an individual on a detailed subject, it also provides an opportunity to observe how the author processes the input paper with his or her background knowledge, and outputs an academic commentary. By comparing the input paper with the academic commentary author's background knowledge (i.e., full-text articles for each of the citations used in the ``output'' paper), the changes in the output paper (i.e., academic commentary) can be visualized by using latent Dirichlet allocation (LDA). LDA use has spread beyond machine learning, and is now a standard analysis tool for researchers in many fields \cite{James2013}. This tool can decompose a collection of full-text materials into several topics.

A third advantage of this scenario is that it is less difficult to obtain the material. Most of the material involved can be accessed through college libraries. To the best of our knowledge, our study is the first to examine a reading and writing relationship using an academic commentary setting. A corpus of academic commentaries on video game addiction was created by collecting full-text articles for each of the citations in the original manuscripts.

\section{Material and methods}

\subsection{Material}
An empirical case we named as ``pros and cons of the concept of video game addiction'' was studied in this experiment. In this case, an original paper \cite{Wood2008} attracted three different academic commentaries: \cite{Alex2008}, \cite{Griffiths2008} and \cite{Turner2008}. According to a response from the original author's own perspective  \cite{Feedback2008}, \cite{Alex2008} focused on the pros of the concept of video game addiction while \cite{Griffiths2008} and \cite{Turner2008} emphasized the cons of it. For a neutral third party (e.g., the readers), it is rather clear to tell that \cite{Alex2008} was actually of the opinion that video game addiction does not exist, whereas \cite{Griffiths2008} and \cite{Turner2008} were essentially against this opinion, and supported the idea that video game addiction does exist.

Table \ref{table:1} shows the demographic information of the material involved.

\begin{table}[htbp]
\caption{Demographic Information of the Material Involved} 
\label{table:1} 
\centering
\begin{threeparttable}
\small
\begin{tabular}{cccc}
\toprule
Paper ( Author* )& Standpoint & Available**  Citations & Total Citations \\
\hline
\cite{Wood2008}(W) & (Original) & Not Considered & 30 \\
\cite{Alex2008}(B) & pros & 2 & 3*** \\
\cite{Griffiths2008}(G) & cons & 16 & 18*** \\
\cite{Turner2008}(T)& cons & 8 & 13**** \\
\bottomrule
\end{tabular}
\begin{tablenotes}
\footnotesize
\item [*] The capital letter in the parentheses of this column is the first letter in the family name of the author, for later reference.
\item [**] Refers to citations with available full texts that can be retrieved completely.
\item [***] The original paper that the commentaries were based on was also included.
\item [****] Two of the citations were different versions of the same manuscript.
\end{tablenotes}
\end{threeparttable}
\end{table}

For the purposes of our study, only academic commentaries with only one author are appropriate, because we are only interested in observing the RRW behaviors of individuals, not groups. Although it is relatively easier to collect materials in our academic commentary scenario, there are still challenges involved in retrieving full text articles for every reference in a commentary manuscript.

In addition to the challenge of simply finding the full texts of the citations, dealing with the various types of references also presents a challenge. Common references include: journal articles, book chapters, institution reports, published proceedings, web pages, entire books, and web portals. Some journal articles are too old and are not accessible on-line for analysis. Some web pages are invalid. Entire books are less meaningful to be included as an individual's background knowledge. Also, most web portals are updated frequently. Due to these reasons, several strategies were developed to retrieve the full texts of the references as completely as possible.
\begin{itemize}
\item For old journal articles:

Their full texts can only be typed into modern electronic documents manually. Even if scanned versions exist, human correction is still needed, even after the help of Optical Character Recognition (OCR) image processing software.

\item For invalid web pages:

Full texts of invalid or expired web pages can typically be collected by searching for their titles or author names over the Internet. Occasionally the content was simply moved to a new website.

\item For web portals:

It is unusual to list web portals as references because its content is updated frequently. If a web portal is listed, it is most likely used to encourage readers to explore the entire portal. In these situations, the web portals were simply discarded.

\item For entire books:

An entire book is cited mostly because the author would like to recommend another different but related domain to readers, similar to web portal citations. However, it can not be simply discarded as a web portal, since its content is quite stable. Nevertheless, the entire book would be too cumbersome to include because of its magnitude and such inclusion would bias the result. So if a much shorter overview can be found to introduce this book, this overview's full text will be regarded as one of the candidates representing the book. It is not difficult to find these candidates (e.g., introductions or formal reviews of the book). These are used in place of an entire book in an effort to avoid the effect of a book's overly large size in the experiment. If the overview from the author of the book is among them, this overview will be used as the substitute of this book for the LDA algorithms. If no such candidates can be found, unfortunately the only option is to discard this citation.
\end{itemize}

The following procedure should only be used when the majority (say over 60\%) of the citations' full text can be retrieved. Otherwise, a new round of finding an academic commentary, like that described in Table 1, is needed.

\subsection{Methods}
The above materials were studied on two levels: \emph{document} and \emph{topic}.

\begin{figure}[htbp]
\flushleft   
\includegraphics[width = 14 cm]{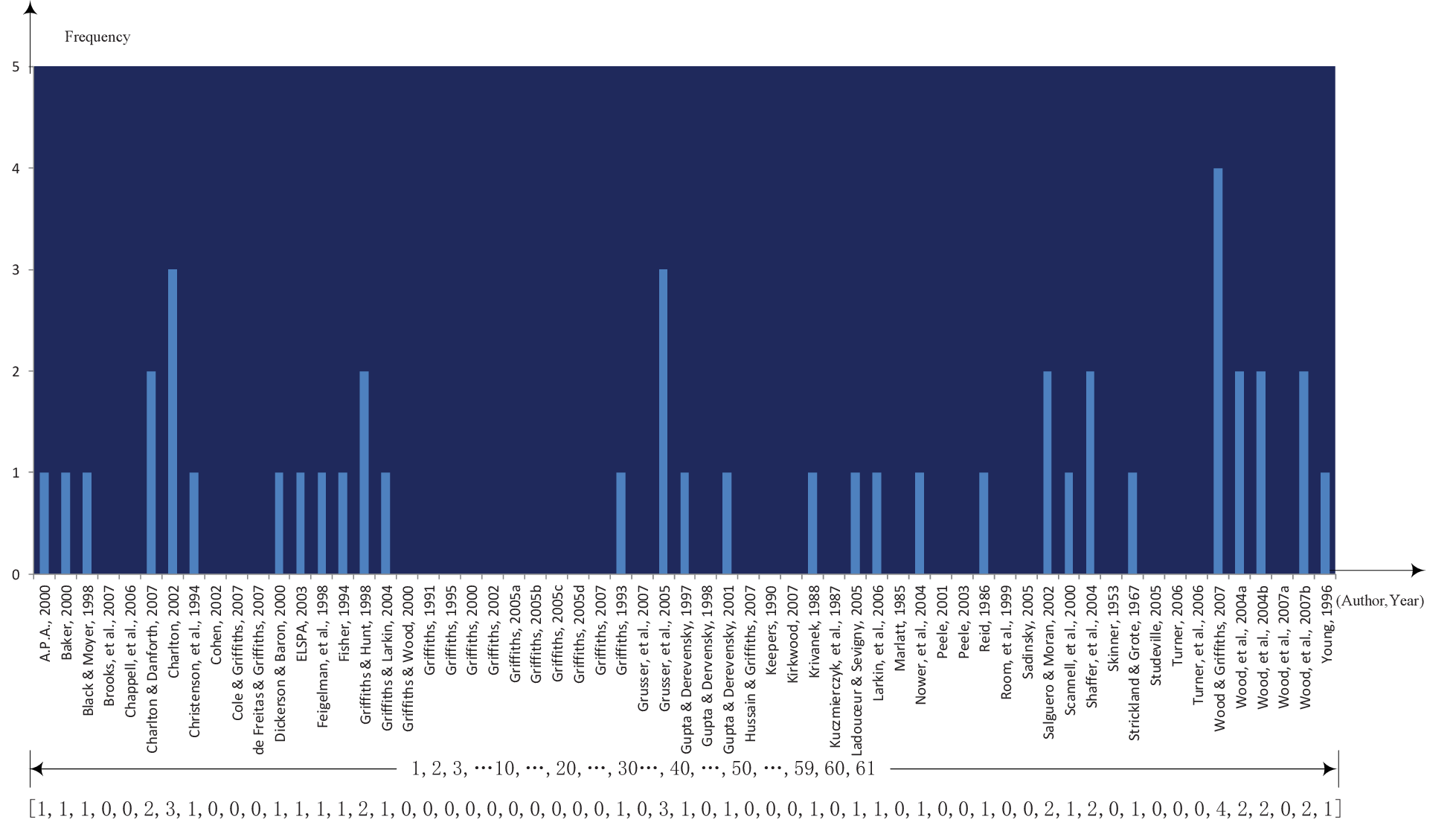}
\caption{Citation Distribution/Vector in \cite{Wood2008}} 
\label{fig:2} 
\end{figure}

\paragraph{Document Level} In this level, a document (more specifically, a commentary) was represented by its citation vector. This vector represented which references were cited, and how many times they were cited in a given document. For example, all together, there were 61 distinct references cited by the papers of authors W, B, G, and T. These indices were then encoded into 1, 2,\dots,61, using the following approach. The earlier the author name appears in the alphabet, and the earlier the citation was, the smaller the index. Fig. \ref{fig:2} illustrates the citation vector of \cite{Wood2008}. The bar height in Fig. 2 represents the citation frequency of a reference \cite{Wood2008}. This citation vector can thus be used to represent the ``parent'' document which generated it.

\paragraph{Topic Level} Although the above citation vector can be used as a mass spectrometer to represent the distribution of some basic features of a certain document, they do not include any semantic information. In order to examine the citations at a deeper level, citation topics were derived semantically. A topic is mathematically defined as a pattern of frequently clustered or co-occurring words within a target corpus. A topic model assumes the words in a corpus cluster into a few unique (perhaps orthogonal \cite{Zhu2009}) topics, with different proportions. In practice, compared to the large number of documents in a corpus, the number of topics is relatively small, otherwise the word distribution a topic represents would become too ambiguous.

One of the most commonly used topic models is LDA. This topic model has become a standard analysis tool in many fields. Recently, a simpler and more efficient stochastic algorithm (SCVB0) than the traditional algorithms was proposed for collapsed variational Bayesian inference for LDA \cite{James2013}. The performance of LDA is determined by several factors. These factors include the number of documents, the length of individual documents and the number of topics\cite{Tang2014}. Unlike KNN-like clustering algorithms, LDA cannot be applied to a collection of words in only a single document; a corpus is needed to derive an interpretable pattern.

The documents in our study were segmented into paragraphs; a decision inspired by the ad hoc heuristics handling with books and tweets mentioned in \cite{Tang2014}.
However, if a paragraph was just a short sentence, it was aggregated into the next paragraph. With this method, we are able to roughly maintain a balance in paragraph length. Mathematically, these separated paragraphs were treated as ``documents'' in \cite{James2013}(the document here was the equivalent of ``collection'' in \cite{James2013}). SCVB0 was then applied directly to the paragraphs. The results of the SCVB0 were combined to represent the topic distribution behind the entire document.

\section{Theory/calculation}

\subsection{A topic process model}
\begin{figure}[htbp]
\flushleft   
\includegraphics[width = 10.4 cm]{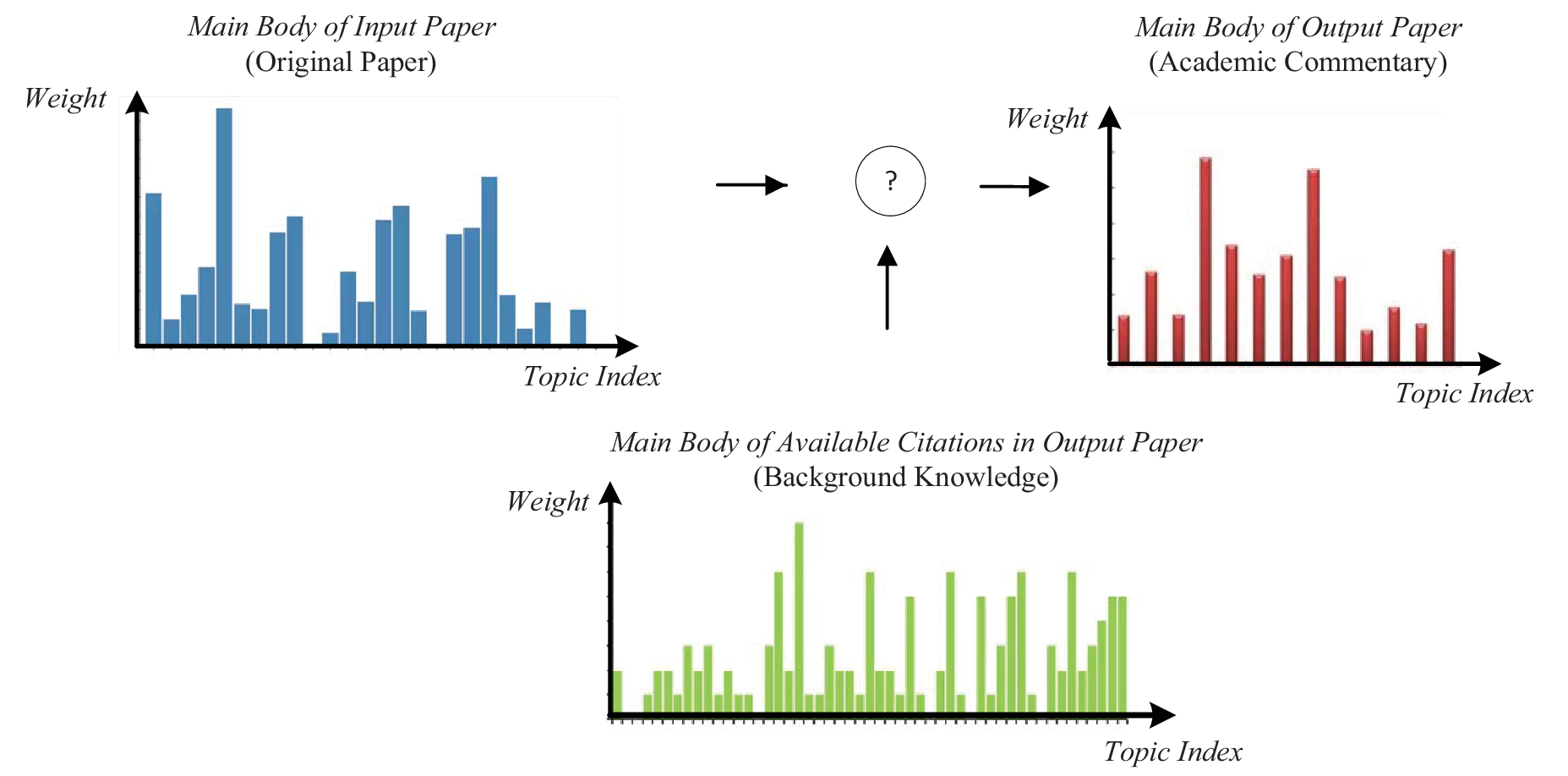}
\caption{Topic Process Model} 
\label{fig:3} 
\end{figure}

This section details a proposed topic process model that was used to gain further insight on RRW in academic commentary publications. As shown in Fig.\ref{fig:3}, the topics in the main body of the output paper are affected by the topics of the input paper, in addition to the topics found in the background knowledge.

In the topic process model, it was assumed that the author processes the input paper with background knowledge in the following way. His or her writing will eventually change the weights of topics based on the background knowledge component ($B_{k}$), with a variation: a function of the product of the input component ($I_{k})$ and ($B_{k}$), as in (\ref{eq:1}). This leads to the output component ($O_{k}$).

\begin{equation}\label{eq:1}
O_{k} = B_{k} + f(B_{k} \cdot I_{k}).
\end{equation}

Eq.(\ref{eq:1}) was partly motivated by the Hebbian learning model, in which the weight of a connection between two neurons is proportional to the product of the signal strengths of the two. This equation is also intuitive, in that the reaction to a given reading material reflects the reader's background knowledge, and the interactions (or correlations) between one's background knowledge and the given reading material. To simplify this, $f(\cdot)$ was defined as a linear function, so (\ref{eq:1}) was transformed into (\ref{eq:2}).

\begin{equation}\label{eq:2}
O_{k} = B_{k} + \gamma_{k}B_{k}I_{k}.
\end{equation}

SCVB0, as introduced in the method subsection, was extended mildly to extract the topic distributions over the paragraphs in the main body of a document. Particularly, the background knowledge was literally represented by the paragraphs in a collection of available citations of the output paper.

Detailed descriptions of these notations are listed in Table \ref{table:2}.

\begin{table}[htbp]
\caption{Summary of Notations} 
\label{table:2} 
\centering
\begin{threeparttable}
\small
\begin{tabular}{cl}
\toprule
Notations & Descriptions \\
\hline
$j$ & The $j$-th paragraph*. \\
$k$ & The $k$-th topic*. \\
$K$ & The number of topics*. \\
$\theta_{jk}$ & The distribution of the $k$-th topic for the $j$-th paragraph*. \\
$\alpha$ & Dirichlet prior parameters for $\theta_{jk}$*. Theoretically, it can be different. \\
         & However, they are usually set equally. \\
$N^{\Theta}_{jk}$ & The expected number of words assigned to the $k$-th topic for the $j$-th \\
                  & paragraph*. \\
$C_{j}$ & The length of the $j$-th paragraph*. \\
$P_I$ & The number of paragraphs in the input (original) paper. \\
$P_O$ & The number of paragraphs in the output paper (commentary). \\
$P_B$ & The number of paragraphs in the citations of the output paper. \\
$\hat{\theta}_{jk}$ & The estimate of $\theta_{jk}$. **\\
$\overline{N}^{\Theta}_{jk}$ & The EM statistics of $N^{\Theta}_{jk}$. *** \\
$I_{k}$ & The sum of $\hat{\theta}_{jk}$ on the $k$-th topic in the input paper. \\
$O_{k}$ & The sum of $\hat{\theta}_{jk}$ on the $k$-th topic in the output paper. \\
$B_{k}$ & The sum of $\hat{\theta}_{jk}$ on the $k$-th topic in the background knowledge. \\
$\gamma_{k}$ & A coefficient of the interactions between $I_{k}$ and $B_{k}$.\\
\bottomrule
\end{tabular}
\begin{tablenotes}
\footnotesize
\item [*] Following the notations and their descriptions in \cite{James2013}.
\item [**] Following the notation and its description in \cite{Cappe2009}.
\item [***] Following the notation and its description in \cite{James2013Long}.
\end{tablenotes}
\end{threeparttable}
\end{table}

\subsection{Calculation}

The paragraphs in the main body of the input (original) paper, the output paper (commentary), and the available citations of the output paper (background knowledge) were put together as a whole collection. Each paragraph in this collection was tagged with a label referencing the proper citation. The paragraphs from the 1st to the $P_I$-th belonged to the input paper. The paragraphs from the ($P_I+1$)-th to the ($P_I+P_B$)-th belonged to the background knowledge. The paragraphs from the ($P_I+P_B+1$)-th to the ($P_I+P_B+P_O$)-th belonged to the output paper.

Suppose SCVB0 was applied to the paragraphs above, then Eq. (12) in \cite{James2013Long} said:

\begin{equation}\label{eq:3}
\hat{\theta}_{jk} = \frac{\overline{N}^{\Theta}_{jk} + \alpha - 1}{C_{j} + K \alpha - K}.
\end{equation}

The estimate of the distribution of the $k$-th topic in the input paper was calculated by summing every $\hat{\theta}_{jk}$ of the paragraphs.
\begin{equation}\label{eq:4}
I_{k} = \sum_{j=1}^{j=P_I}\hat{\theta}_{jk} = \sum_{j=1}^{j=P_I} \frac{\overline{N}^{\Theta}_{jk} + \alpha - 1}{C_{j} + K \alpha - K}.
\end{equation}

According to the paragraph aggregation / document segmentation method presented in the subsection 2.2, it was assumed that each length of paragraphs is approximately the same. That means: $C_j \approx C_I (j = 1,2,..., P_I)$. Then,

\begin{equation}\label{eq:5}
I_{k} \approx \frac{1}{C_I + K(\alpha - 1)}\sum_{j=1}^{j=P_I}[{\overline{N}^{\Theta}_{jk} + (\alpha - 1)}].
\end{equation}

A notation for the mean function, $\textbf{mean}(\overline{N}^{\Theta}_{jk})_{I}$,
was introduced:
\begin{equation}\label{eq:6}
\sum_{j=1}^{j=P_I}(\overline{N}^{\Theta}_{jk}) = P_I \cdot \textbf{mean}(\overline{N}^{\Theta}_{jk})_{I}.
\end{equation}

After substituting Eq. (\ref{eq:6}) to Eq. (\ref{eq:5}), Eq. (\ref{eq:5}) was transformed into:

\begin{equation}\label{eq:7}
I_{k} \approx \frac{P_I}{C_I}[\frac{1}{\frac{1}{\alpha-1}+\frac{K}{C_I}} + \frac{\textbf{mean}(\overline{N}^{\Theta}_{jk})_{I}}{1+\frac{K}{C_I}(\alpha-1)}].
\end{equation}

In a similar way, Eq. (\ref{eq:8}) and Eq. (\ref{eq:9}) were obtained:

\begin{equation}\label{eq:8}
B_{k} \approx \frac{P_B}{C_B}[\frac{1}{\frac{1}{\alpha-1}+\frac{K}{C_B}} + \frac{\textbf{mean}(\overline{N}^{\Theta}_{jk})_{B}}{1+\frac{K}{C_B}(\alpha-1)}],
\end{equation}

\begin{equation}\label{eq:9}
O_{k} \approx \frac{P_O}{C_O}[\frac{1}{\frac{1}{\alpha-1}+\frac{K}{C_O}} + \frac{\textbf{mean}(\overline{N}^{\Theta}_{jk})_{O}}{1+\frac{K}{C_O}(\alpha-1)}],
\end{equation}

where $C_j \approx C_B (j = P_I+1,...,P_I+P_B)$ and $C_j \approx C_O (j = P_I+P_B+1,...,P_I+P_B+P_O)$. $\textbf{mean}(\overline{N}^{\Theta}_{jk})_{B}$ and $\textbf{mean}(\overline{N}^{\Theta}_{jk})_{O}$ were the equivalents of $\textbf{mean}(\overline{N}^{\Theta}_{jk})_{I}$, in the background knowledge and the
output paper respectively. Briefly speaking, they were closely related to the average density of words that the author spent on the $k$-th topic in a certain paragraph.

By
applying SCVB0, $I_k$ was obtained by Eq. (\ref{eq:4}), so were $B_k$ and $O_k$, in a similar way. And then $\gamma_k$ can be solved by Eq. (\ref{eq:2}).
This parameter shows to what extent the input and background knowledge interact with each other, and eventually shapes the output.

\section{Results}

The three authors in Table \ref{table:1} actually represented three sub-cases in our case. Each of their commentaries was a complete scenario presented by the topic process model in Fig. \ref{fig:3}. For each scenario, the demographic information of the collections are listed in Table \ref{table:3}.

\begin{table}[htbp]
\caption{Demographic Information of the Scenarios} 
\label{table:3} 
\centering
\begin{threeparttable}
\small
\begin{tabular}{ccccc}
\toprule
Author & Articles* & Paragraphs & Topics** & Tokens***\\
\hline
B & 4 & 131 & 5 & 25685 \\
G & 18 & 431 & 9 & 58173 \\
T & 10 & 391 & 7 & 59303\\
\bottomrule
\end{tabular}
\begin{tablenotes}
\footnotesize
\item [*] Articles included the input and output papers.
\item [**] They were set after a few trials according to the principles summarized by \cite{Tang2014}.
\item [***] If a word occurred twice, two tokens were included.
\end{tablenotes}
\end{threeparttable}
\end{table}

\subsection{Document Level}

\begin{figure}[htbp]
\centering   
\includegraphics[width = 10.4 cm]{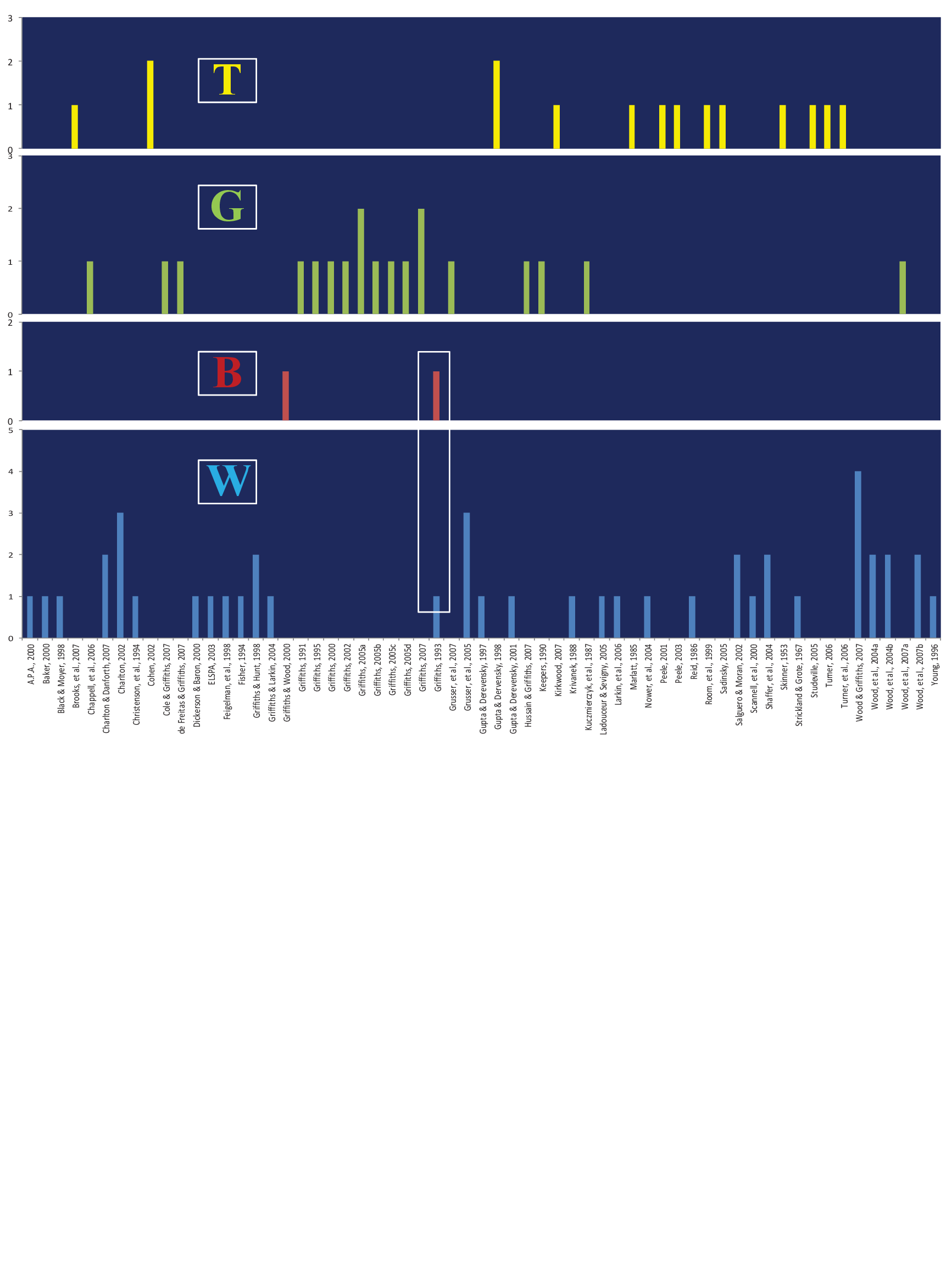}
\caption{Citation Vectors} 
\label{fig:4} 
\end{figure}

The citation vectors of the above sub-cases are visualized in Fig. \ref{fig:4}, including the original (target) paper itself.

It was found that there was only one overlap between these citation vectors. This overlap is highlighted by a white rectangle. Although this overlap was particularly rare, it was hardly coincidental, as evidenced by the proportion of frequencies. Half of the citations (though there were only two altogether) in paper B overlapped with paper W which had 30 citations, while 17 citations in paper G and 13 citations in paper T overlapped with none of the citations in paper W. The explanation for this phenomenon is as follows: B focused on the pros of paper W, so the probability of overlaps was much higher than the other two sub-cases. This contrast illustrates that scientific researchers tend to cite the most direct and solid evidence to support their opinions, and some of the classic or more well-known evidence can be traced to the same reference. Alternatively, G and W co-authored several papers in the related area, but within this theme, there was zero overlap between their citations. This showed the finding above again, but from a different aspect: when researchers oppose each other on a topic, the references they cite are much less likely to overlap.

We can borrow a term commonly used in Physics, ``resonance'', to describe the above phenomenon observed in our study. When two scientific researchers agree with each other on a specific theme, something must trigger a strong resonance in their minds. In other words, there must be something significant that co-exists in their background knowledge. We argue the overlapping references observed in our study provides support for this idea.

\subsection{Topic Level}
\begin{figure}[htbp] \centering
\subfigure[Author B.] { \label{fig:5a}
\includegraphics[width=7.2cm]{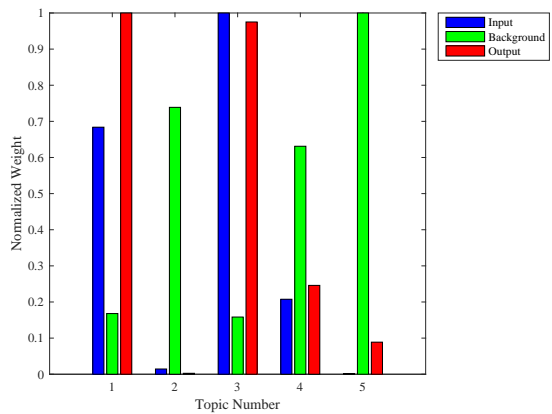}
}\\
\subfigure[Author G.] { \label{fig:5b}
\includegraphics[width=7.2cm]{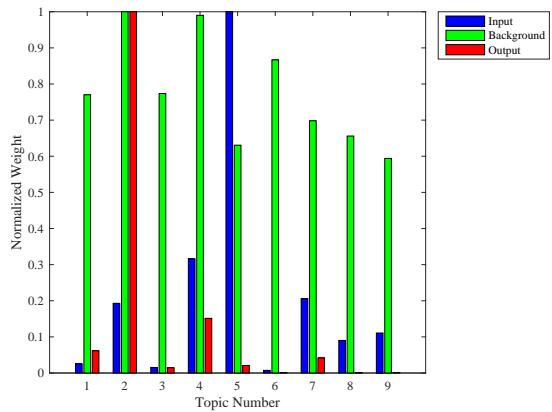}
}\\
\subfigure[Author T.] { \label{fig:5c}
\includegraphics[width=7.2cm]{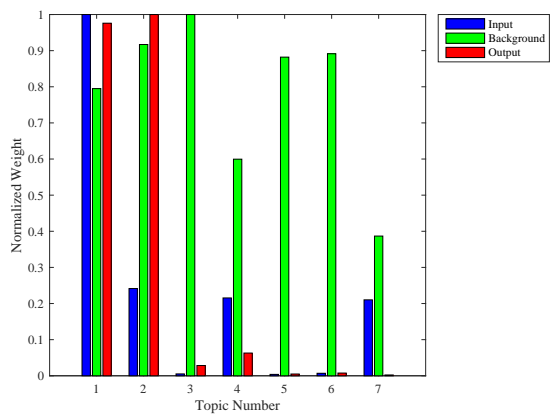}
}
\caption{Topic Process Model Entailed in Three Different Commentaries}
\label{fig:5}
\end{figure}

The weighted word density spent on a certain topic in one of the commentaries is presented in Fig.\ref{fig:5}. The implementation of SCVB0 used a variable named ``weight'' to record the topic distribution. Fig.\ref{fig:5} inherited this notation, but was provided with a different meaning, the sum of weight, which was consistent with Eq.(\ref{eq:4}). The height of each bar in Fig.\ref{fig:5} was normalized by the maximum weight among the bars with the same color. The maximum height of a bar is set at 1. These three sub-figures illustrate that both input and background knowledge modulate the output. The generated topics along with the top ten words of each scenario are listed in Table \ref{table:6}. The underlined word was selected as a potential representative for the topic. Two relationships (input vs. output, and background knowledge vs. output) were analyzed.

\begin{table}[htbp] 
\tiny
\hspace{1.2cm}
\subtable[\footnotesize{Case: Author B.}]  { \label{table:6a}
\begin{tabular}{lllll}
\toprule
Topic 1 & Topic 2 & Topic 3 & Topic 4 & Topic 5\\
\hline
video & gambling & video & gambling & gambling\\
playing & machine & game & internet & \underline{characteristics}\\
time & machines & playing & problem & machines\\
game & play & games & factors & structural\\
addiction & slot & problems & gaming & machine\\
games & \underline{adolescents} & people & many & fruit\\
behaviour & playing & addiction & \underline{money} & gamblers\\
\underline{excessive} & players & behaviour & use & winning\\
gaming & fruit & problem & research & pay\\
play & lottery & \underline{case} & slot & player\\
\bottomrule
\end{tabular}
}\\

\tiny
\hspace{-0.9cm}
\subfigure[\footnotesize{Case: Author T.}] { \label{table:6c}
\begin{tabular}{lllllll}
\toprule
Topic 1 & Topic 2 & Topic 3 & Topic 4 & Topic 5 & Topic 6 & Topic 7\\
\hline
video & gambling & gambling & gambling & gambling & gambling & casino\\
game & gamblers & \underline{niagara} & \underline{addiction} & \underline{strategy} & gamblers & community\\
\underline{relapse} & \underline{pathological} & falls & alcohol & treatment & \underline{reported} & casinos\\
behaviour & problem & casino & addictive & problem-gambling & pathological & gambling\\
games & people & problems & treatment & problem & females & time\\
playing & win & respondents & problem & commission & males & also\\
treatment & big & data & model & gaming & grade & problems\\
time & variables & items & drug & ontario & age & \underline{expectations}\\
play & events & pathological & coping & services & gamble & increase\\
client & non-problem & sample & individuals & research & found & people\\
\bottomrule
\end{tabular}
}\\

\tiny
\hspace{-1.27cm}
\subtable[\footnotesize{Case: Author G.}]  { \label{table:6b}
\begin{tabular}{lllllllll}
\toprule
Topic 1 & Topic 2 & Topic 3 & Topic 4 & Topic 5 & Topic 6 & Topic 7 &  Topic 8 & Topic 9\\
\hline
players & addiction & online & video & time & online & gambling & video & video\\
playing & videogame & friends & games & playing & gaming & playing & games & games\\
play & videogames & male & game & game & study & \underline{poker} & game & playing\\
week & addictive & internet & machine & video & \underline{mmorpgs} & online & children & game\\
per & excessive & female & machines & games & players & gamblers & playing & players\\
played & \underline{playing} & gender & playing & play & gamers & problem & used & \underline{treatment}\\
game & games & \underline{extract} & \underline{fruit} & life & everquest & likely & social & effects\\
gamers & addictions & gamers & amusement & \underline{behaviour} & game & problems & \underline{agrressive} & people\\
\underline{hours} & many & real & use & computer & data & money & skills & children\\
reported & research & life & research & parents & participants & play & reported & control\\
\bottomrule
\end{tabular}
}
\caption{\footnotesize{Top Ten Words in Each Topic}}
\label{table:6}
\end{table}

\paragraph{On the relationship between input and output} The fluctuations of input and output appear to be in a similar pattern in Fig.\ref{fig:5a}. Except for Topic 5, almost every weight of a topic in the input and output paper rose and fell at the same phase. The pattern in this sub-figure was much more obvious than in the other two sub-figures. Considering Author B focused on the pros of the original paper, this phenomenon could be naturally attributed to the fact that both Author B and W held the same opinion.

The correlation coefficients between the weights of topics of input and the ones of outputs in these three sub-cases are listed in Table \ref{table:4}. As seen in this table, the correlation coefficient of G was quite close to zero. This implies that the output of G was nearly irrelevant to the input to him. In other words, G's opposition to W's opinion was the strongest among the three scholars. To support this inference, consider the following circumstantial evidence. First, G and W co-wrote several papers in related areas, however, they did not do so on this debate theme. Secondly, the lines in the input and output papers revealed their disagreements.

\begin{table}[htbp]
\caption{Correlation Coefficients between Input and Output} 
\label{table:4} 
\centering
\small
\begin{tabular}{cc}
\toprule
Author & Correlation Coefficient\\
\hline
B & 0.9600\\
G & -0.0124 \\
T & 0.7326 \\
\bottomrule
\end{tabular}
\end{table}

\paragraph{On the relationship between background knowledge and output} The observations became more complicated when taking Fig.\ref{fig:5b} and Fig.\ref{fig:5c} into account. A similar same-phase fluctuating pattern of input and output was observed in Topic 3 of Fig.\ref{fig:5b} and Topic 1 of Fig.\ref{fig:5c}. However, more fluctuations of input and output exhibited opposite-phases, especially Topic 2 and 5 of Fig.\ref{fig:5b}, and Topic 2 and 7 of Fig.\ref{fig:5c}. For these opposite-phases fluctuations of input and output, the background knowledge appears to be a more significant factor than the input, when shaping the output.

Besides the above topics, the background knowledge also dominates in some other topics, e.g., Topic 4 in Fig.\ref{fig:5b}. In this sub-figure, Topic 4 has the second greatest normalized weight among the topics, \emph{both} in the background knowledge and output paper.

The correlation coefficients between the weights of topics of background knowledge and the weights of outputs in these 3 sub-cases are listed in Table \ref{table:5}. A surprising and interesting result was found in the negative correlation coefficient of B ( $r = -0.9446$). B's correlation coefficient implies that B used a very different strategy when applying his background knowledge to compose his commentary. He might be citing references to help organize the manuscript as seen in the emphasis on the topics with less focus on the input. He could be citing references that he opposes. He could also be using a combination of both strategies. The possible strategies warrant further exploration in the future.

\begin{table}[htbp]
\caption{Correlation Coefficients between Background Knowledge and Output} 
\label{table:5} 
\centering
\small
\begin{tabular}{cc}
\toprule
Author & Correlation Coefficient\\
\hline
B & -0.9446 \\
G & 0.6437 \\
T & 0.2353 \\
\bottomrule
\end{tabular}
\end{table}

\section{Discussion}

The above results visualize how the input paper and background knowledge take part in shaping the output paper. The parameter $\gamma$ in Eq.(\ref{eq:2}) reflects the strategy used to process and combine the input he or she reads with his or her background knowledge when writing the output. This parameter may only work for this specific input, and may not generalize to other cases (e.g., posting on Twitter). However, if the author's other commentaries can be collected, his or her own style of a topic processing model could be induced using the method described in this paper.

Many, if not all, scientific problems have multiple sides to explore. When a scientific debate occurs, scientists, even colleagues, may find themselves on different sides. One may emphasize a specific side (topic), where a different scientist may focus on another. This ``emphasizing'' or ``focusing'' goal can drive the author to either consciously or unconsciously allocate (i.e., organize) different amounts of words on different topics. Scholars of a similar view or mindset on a specific problem may adopt the same strategies in using technical terms. Scholars of different views or mindsets appear to allocate different proportions of technical terms. This explanation may encounter one exception: someone uses the same expression but with a different or opposite meaning. However, authors may be discouraged to use figures of speech, or words that can be interpreted in multiple ways, because it is an additional source of confusion.

This finding is consistent with, and extends on, the findings in \cite{R2013}. \cite{R2013} indicates that students' reading of multiple documents is influenced by perspective instructions (i.e., themes or topics), which, in turn, can help students be more discriminative when deciding between more and less trustworthy documents. Specifically, participants were assigned different stances on a controversial topic, prior to collecting evidence to support their stance. The perspective instructions in their work could be viewed as a type of ``input''.

Furthermore, our study provides support for the idea that background knowledge can play a significant role in shaping the output paper. \cite{Anmarkrud2013} found that readers discriminate between more- and less-relevant information while they read. Readers were more likely to build connections between more-relevant information and other texts, than they were with less-relevant information. More- and less-relevant information here can be seen as the background knowledge. Meanwhile, \cite{Bos2015} suggests that creative writing requires a similar representational process as reading comprehension. Our study not only replicates their findings, but also provides a visualization of the same-phase or opposite-phase fluctuations among background knowledge and the output paper.

In summary, our empirical study shows that the output paper is modulated by both the input paper and the background knowledge. The results of this study also has implications on education and learning science, in that there appears to exist a latent ``resonance'' hidden between the learner's background knowledge and the reading material. The results also indicate that a learner's background knowledge can be estimated when equipped with information about what the learner has read and written. With this estimated background knowledge, educational material can be created to trigger the latent ``resonance'' hidden in the background knowledge.

\section{Conclusion}

The relationship between reading and writing was studied under a special scenario: academic commentary. This scenario was chosen to explore RRW because it provided a unique advantage in reducing the difficulty to obtain accurate background knowledge, and also because it was believed that the implications from this setting could shed light on RRW behaviors in general.

The academic commentary materials for this study were divided into three parts: the target article that the commentary comments on, the commentary article itself, and the commentary article's citations. Each part corresponded to the input, output, and the background knowledge of the same author, respectively. SCVB0, a simpler and more efficient version of LDA algorithm, was extended to visualize the relationship among the three parts, under the framework of a proposed topic process model.

The observed ``resonance'' implied that the output paper is influenced both by input and background knowledge. More specifically, it was found that the fluctuations of input, output, and background knowledge can exhibit a same-phase or an opposite-phase. Future work should involve two different directions. First, sample scales should be increased to observe how the parameter changes. More academic commentaries should be collected and tested, in addition to testing the topic process model in more controllable groups (e.g., elementary students). The ultimate goal is to see if the parameter $\gamma$ can be related to the psychometrics results of these participants. Secondly, future work should move forward into the word level, similar to \cite{Lee2015}.

\section*{Acknowledgement}
The authors would like to extend their thanks to Prof. Mark Steyvers for his generous recommendation on SCVB0 when they met in Seoul National University, Korea, at APCCBS, Oct., 2014. The authors would also like to extend their thanks to Prof. Dietrich Albert for his valuable comments on the experiments when he paid a visit in Wuhan, China, Nov., 2015.

The work presented was financially supported by the National High-Tech R\&D
Program of China (863 Program)(No. 2014AA015103) and the Humanity and Social Science Youth foundation of Ministry of Education of China (No. 2015YJC880088).


\bibliography{mybibfile}

\begin{thebibliography}{10}
\expandafter\ifx\csname url\endcsname\relax
  \def\url#1{\texttt{#1}}\fi
\expandafter\ifx\csname urlprefix\endcsname\relax\def\urlprefix{URL }\fi
\expandafter\ifx\csname href\endcsname\relax
  \def\href#1#2{#2} \def\path#1{#1}\fi

\bibitem{Harl2013}
A.~L. Harl, A historical and theoretical review of the literature: Reading and
  writing connections, Reconnecting Reading and Writing, publisher: The WAC
  Clearinghouse (2013) 26--54.

\bibitem{Abbott2010}
R.~D. Abbott, {V. W. Berninger, M. Fayol}, Longitudinal relationships of levels
  of language in writing and between writing and reading in grades 1 to 7,
  Journal of Educational Psychology 102(2) (2010) 281--298.

\bibitem{Mateos2011}
M.~Mateos, {I. Cuevas, E. Martín, A. Martín, G. Echeita, M. Luna}, Reading to
  write an argumentation: the role of epistemological, reading and writing
  beliefs, Journal of Research in Reading 34 (2011) 281–--297.

\bibitem{James2013}
J.~Foulds, L.~{Boyles, C. DuBois, P. Smyth and M. Welling}, Stochastic
  collapsed variational bayesian inference for latent dirichlet allocation, in:
  the 19th ACM SIGKDD international conference on Knowledge discovery and data
  mining, 2013, pp. 446--454.

\bibitem{Wood2008}
R.~T.~A. Wood, Problems with the concept of video game ``addiction'': some case
  study examples, International Journal of Mental Health and Addiction 6~(2)
  (2008) 169--178.

\bibitem{Alex2008}
A.~Blaszczynski, Commentary: a response to ``problems with the concept of video
  game ``addiction'': some case study examples'', International Journal of
  Mental Health and Addiction 6~(2) (2008) 179--181.

\bibitem{Griffiths2008}
M.~D. Griffiths, Videogame addiction: further thoughts and observations,
  International Journal of Mental Health and Addiction 6~(2) (2008) 182--185.

\bibitem{Turner2008}
N.~E. Turner, A comment on ``problems with the concept of video game
  ‘addiction’: some case study examples'', International Journal of Mental
  Health and Addiction 6~(2) (2008) 186--190.

\bibitem{Feedback2008}
R.~T.~A. Wood, A response to blaszczynski, griffiths and turners’ comments on
  the paper ``problems with the concept of video game ‘addiction’: some
  case study examples'', International Journal of Mental Health and Addiction
  6~(2) (2008) 191--193.

\bibitem{Zhu2009}
D.~Andrzejewski, {X. Zhu}, Latent dirichlet allocation with topic-in-set
  knowledge, in: North American Chapter of the Association for Computational
  Linguistics Workshop on Semi-supervised Learning for NLP (NAACL-SSLNLP 2009),
  2009.

\bibitem{Tang2014}
J.~Tang, Z.~{Meng, X. Nguyen, Q. Mei and M. Zhang}, Understanding the limiting
  factors of topic modeling via posterior contraction analysis, in: the 31st
  International Conference on Machine Learning, JMLR Workshop and Conference
  Proceedings, Vol.~32, 2014.

\bibitem{Cappe2009}
O.~Cappé, E.~Moulines, On-line expectation–-maximization algorithm for
  latent data models, Journal of the Royal Statistical Society: Series
  B(Statistical Methodology) 71~(3) (2009) 593--613.

\bibitem{James2013Long}
J.~Foulds, L.~{Boyles, C. DuBois, P. Smyth and M. Welling}, Stochastic
  collapsed variational bayesian inference for latent dirichlet allocation,
  arXiv:1305.2452v1(14 May, 2013).

\bibitem{R2013}
R.~Cerdána, {M. C. Marínb and C. Candel}, The role of perspective on
  students’ use of multiple documents to solve an open ended task,
  Psicología Educativa 19 (2013) 89--94.

\bibitem{Anmarkrud2013}
O.~Anmarkrud, {M. T. McCrudden, I. Bråten , H. I. Strømsø}, Task-oriented
  reading of multiple documents: online comprehension processes and offline
  products, Instructional Science 41 (2013) 873--894.

\bibitem{Bos2015}
L.~T. Bos, {B. B. de Koning, F. v. Wesel, A. M. Boonstra, M. v. der Schoot},
  What can measures of text comprehension tell us about creative text
  production?, Read and Write 28 (2015) 829--849.

\bibitem{Lee2015}
S.~Lee, { Y. Park, W. C. Yoon}, Burst analysis for automatic concept map
  creation with a single document, Expert Systems With Applications 42 (2015)
  8817–--8829.

\end{thebibliography}

\end{document}